\documentclass[a4paper,11pt]{amsart}
\begin{document}

\hyphenation{gra-vi-ta-tio-nal re-la-ti-vi-ty Gaus-sian
re-fe-ren-ce re-la-ti-ve gra-vi-ta-tion Schwarz-schild
ac-cor-dingly gra-vi-ta-tio-nal-ly re-la-ti-vi-stic pro-du-cing
de-ri-va-ti-ve ge-ne-ral ex-pli-citly des-cri-bed ma-the-ma-ti-cal
de-si-gnan-do-si coe-ren-za pro-blem gra-vi-ta-ting geo-de-sic
per-ga-mon cos-mo-lo-gi-cal gra-vity cor-res-pon-ding
de-fi-ni-tion phy-si-ka-li-schen ma-the-ma-ti-sches ge-ra-de
Sze-keres con-si-de-red tra-vel-ling ma-ni-fold re-fe-ren-ces
geo-me-tri-cal in-su-pe-rable}

\title[Attraction and repulsion \emph{etc.}]
{{\bf Attraction and repulsion in spacetime\\of an electrically
charged mass-point}}

\author[Angelo Loinger]{Angelo Loinger}
\address{A.L. -- Dipartimento di Fisica, Universit\`a di Milano, Via
Celoria, 16 - 20133 Milano (Italy)}
\author[Tiziana Marsico]{Tiziana Marsico}
%\date{}
\address{T.M. -- Liceo Classico ``G. Berchet'', Via della Commenda, 26 - 20122 Milano (Italy)}
\email{angelo.loinger@mi.infn.it} \email{martiz64@libero.it}
%\thanks{}

\vskip0.50cm

\begin{abstract}
By virtue of Hilbert repulsive effect, the most external singular
surface of Rei\ss ner-Weyl-Nordstr\"om metric represents an
insuperable barrier for the arriving neutral particles and light
rays.
\end{abstract}

\maketitle
%%\begin{equation} \label{eq:sevenprime}
%%    \ddot{\Re} + \f\textbf{5}. External
%% {\kappa}{6}\Re \rho=0 , \tag{7'}
%% \end{equation}
%% ``mechanisms'' \textrm{d} \`a
%% \cite{1}
%% eqs.(\ref{eq:six})
%% Schwarzschild


\normalsize

\vskip1.20cm \noindent \textbf{1.} -- \emph{The metric}. -- If
$\gamma\equiv 1- (2m/r)+(q^{2}/r^{2})$, we have (see
\cite{1}$\div$\cite{7}):

\begin{equation} \label{eq:one}
\textrm{d}s^{2} = \gamma \, \textrm{d}t^{2} - \gamma^{-1}
\textrm{d}r^{2} -r^{2}\textrm{d}\omega^{2} \quad; \quad (G=c=1)
\quad,
\end{equation}

where: $\textrm{d}\omega^{2} \equiv
\textrm{d}\vartheta^{2}+\sin^{2}\vartheta \,
\textrm{d}\varphi^{2}$; $q^{2}\equiv 4\pi\varepsilon^{2}$; and
$4\pi\varepsilon$ is the electric charge of the gravitating
point-mass $m$; $\gamma (r=0) = +\infty$; $\gamma (r=\infty) = 1$.

\par Two cases: $m^{2}< q^{2}$, in particular $m=0$; and $m^{2}\geq
q^{2}$. (For the electron we have $m^{2}< q^{2}$).

\par If $m^{2}< q^{2}$, $\gamma(r)$ is everywhere positive, with a
minimal value at $r=q^{2}/m$:
$\gamma_{\textrm{min}}=1-(m^{2}/q^{2})$; if $m=0$,
$\gamma_{\textrm{min}}=1$.

\par If $m^{2}= q^{2}$, $\gamma(r)=\left[1-(m/r)\right]^{2}$;
$\gamma_{\textrm{min}}=\gamma(r=m)=0$;
$\gamma_{\textrm{max}}=\gamma(r=\infty)=1$.

\par If $m^{2}> q^{2}$, we can write:

\begin{equation} \label{eq:two}
r^{2}\,\gamma(r) = \left[r-m-(m^{2}-q^{2})^{1/2}\right] \, \cdot
\, \left[r-m+(m^{2}-q^{2})^{1/2}\right] \quad ;
\end{equation}

let us put: $r_{1}\equiv m+(m^{2}-q^{2})^{1/2}$ and $r_{2}\equiv
m-(m^{2}-q^{2})^{1/2}$. On the spherical surfaces $r=r_{1}$ and
$r=r_{2}$, we have $\gamma(r_{1,2})=0$. It is easy to see that
when $r_{2}<r<r_{1}$, $\gamma(r)<0$, but $\gamma(r<r_{2})>0$.

\vskip1.20cm \noindent \textbf{2.} -- \emph{Radial geodesics of
light-rays}. -- If $\textrm{d}s^{2}=0=\textrm{d}\omega^{2}$, we
have $\gamma \, \textrm{d}t^{2}= \gamma^{-1} \textrm{d}r^{2}$,
from which:

\begin{equation} \label{eq:three}
\left(\frac{\textrm{d}r}{\textrm{d}t}\right)^{2} \equiv
\dot{r}^{2} = \frac{(r^{2}-2\,m\,r+q^{2})^{2}}{r^{4}} \quad;
\end{equation}

$\left[\ \dot{r}^{2}\right]_{r=\infty}=1$; $\left[\
\dot{r}^{2}\right]_{r=0}=\infty$. For $m^{2}>q^{2}$, we have
$\left[\ \dot{r}^{2}\right]_{r=r_{1,2}}=0$; and for $m^{2}=q^{2}$,
$\left[\ \dot{r}^{2}\right]_{r=m}=0$.

\par The integration of \cite{3} gives:

\begin{equation} \label{eq:four}
t + \textrm{const} = \pm \int \frac{r^{2}\,\textrm{d}r}{r^{2}-2\,
m\,r+q^{2}} \quad;
\end{equation}

thus, when $m^{2}>q^{2}$:

\begin{eqnarray} \label{eq:five}
\pm \, (t + \textrm{const}) & = & r+ m\,\ln
|\,r^{2}-2\,m\,r+q^{2}| +
\nonumber\\
& & {} %%
+ \frac{2\,m^{2}-q^{2}}{2\,(m^{2}-q^{2})^{1/2}} \,\,  \ln \left|
\frac{r-m-(m^{2}-q^{2})^{1/2}}{r-m+(m^{2}-q^{2})^{1/2}}\right|\quad;
\end{eqnarray}

this means that a light-ray which starts from an $r>r_{1}$,
arrives at $r=r_{1}$ after an \emph{infinite} time interval (and
with a velocity equal to zero): we have a Hilbertian gravitational
\emph{repulsion}.

\par When $m^{2}=q^{2}$:

\begin{equation} \label{eq:six}
\pm \, (t + \textrm{const}) =  r- m +2\,m \,\ln |\,r-m|
-m^{2}/(r-m) \quad;
\end{equation}

for $r=m$, $\pm \, (t + \textrm{const}) = \infty$: Hilbertian
repulsion, as far $m^{2}>q^{2}$ (and a velocity $\left[\
\dot{r}\right]_{r=m}=0$).

\par When $m^{2}<q^{2}$, we have:

\begin{eqnarray} \label{eq:seven}
\pm \, (t + \textrm{const}) & = & r+ m\,\ln
|\,r^{2}-2\,m\,r+q^{2}| +
\nonumber\\
& & {} %%
+ \frac{2\,m^{2}-q^{2}}{(q^{2}-m^{2})^{1/2}} \, \cdot \, \arctan
\left[ \frac{r-m}{(q^{2}-m^{2})^{1/2}}\right]\quad;
\end{eqnarray}

a light-ray arrives at $r=0$ in a finite time: gravitational
attraction.

\par When $m=0$:

\begin{equation} \label{eq:eight}
\pm \, (t + \textrm{const}) =  \int
\frac{r^{2}\textrm{d}r}{r^{2}+q^{2}} = r- q \, \arctan \, (r/q)
\quad;
\end{equation}

attraction, as for $m^{2}<q^{2}$.

\par Let us compute the acceleration $\ddot{r}$:

\begin{equation} \label{eq:nine}
\pm \, \ddot{r} = 2\, r^{-5} \, (r^{2}-2\,m\,r+q^{2}) \,
(m\,r-q^{2})\quad;
\end{equation}

of course, $\left[\ddot{r}\right]_{r=\infty}=0$. We have:
\vskip0.20cm \noindent

for $m^{2}>q^{2}$: $\ddot{r}=0$ at $r=r_{1}$, repulsion;
$(\ddot{r}=\infty$ at $r=0)$; \

for $m^{2}=q^{2}$: $\ddot{r}=0$ at $r=m$, repulsion;
$(\ddot{r}=\infty$ at $r=0)$; \

for $m^{2}<q^{2}$: $\ddot{r}=\infty$ at $r=0$, attraction;
$\ddot{r}=0$ at $r=q^{2}/m$; \

for $m=0$: $\ddot{r}=\infty$ at $r=0$, attraction.

\par \vskip0.20cm When $m^{2}\geq q^{2}$ a light-ray which arrives at $r=r_{1}$
starting from an $r>r_{1}$ cannot overcome the barrier $r=r_{1}$:
indeed, velocity $\dot{r}$ and acceleration $\ddot{r}$ are zero at
$r=r_{1}$. (On the contrary, gravitational attraction predominates
in the interval $0\leq r < r_{2}$).

\vskip1.20cm \noindent \textbf{3.} -- \emph{Radial geodesics in
general}. -- We have the following first integral:

\begin{equation} \label{eq:ten}
\dot{r}^{2} = \gamma^{2} (1-|A|\,\gamma)  \quad,
\end{equation}

where $A<0$ for test-particles, and $A=0$ for light-rays. We get
from eq. (\ref{eq:ten}) that

\begin{equation} \label{eq:eleven}
\pm \, \ddot{r} = \gamma \, r^{-2} (m-q^{2}r^{-1}) \, (2-3|A|\,
\gamma) \quad.
\end{equation}

\emph{The case} $m^{2}\geq q^{2}$ \emph{for} $r>r_{1}\equiv
m+(m^{2}-q^{2})^{1/2}$ \emph{is interesting}. When $\ddot{r}=0$,
attraction and repulsion \emph{balance} each other. Putting
$\ddot{r}=0$ in eq. (\ref{eq:eleven}), we obtain

\begin{equation} \label{eq:twelve}
|A| \, \gamma = \frac{2}{3} \quad,
\end{equation}

and substituting eq. (\ref{eq:twelve}) in eq. (\ref{eq:ten}):

\begin{equation} \label{eq:thirteen}
\dot{r}^{2}_{b} = \frac{1}{3} \, \gamma^{2} \quad;
\end{equation}

a motion described by the balance velocity $\dot{r}_{b}$ has a
zero acceleration. Where $\dot{r}^{2}<\dot{r}^{2}_{b}$, there is
attraction; where $\dot{r}^{2}>\dot{r}^{2}_{b}$ there is
\emph{repulsion}.

\par At $r=r_{1}$, velocity $\dot{r}$ and acceleration $\ddot{r}$
are equal to zero: \emph{no} particle, \emph{no} light-ray can
overcome this barrier. Remark that for the light-rays there is
repulsion on the \emph{whole} trajectory $(m^{2}\geq q^{2})$.

\par The study of the \emph{circular} geodesics confirms the
existence of the Hilbertian repulsion: as for the Schwarzschild
manifold (see \cite{8}), there is a \emph{minimal} value for the
radial coordinate $r$  of the test-.particles, and a \emph{unique}
value for the $r$ of light-rays $(m^{2}\geq q^{2})$.

\par When $q^{2}>m^{2}$, there is no gravitational repulsion.

\par A remark. If the term $q^{2}/r^{2}$ of $\gamma(r)$, which is
originated by the mass-energy of the electrostatic field
$\varepsilon/r^{2}$, were endowed with a Newtonian nature, it
would have a repulsive action. On the contrary, it has in GR an
attractive effect, as it is particularly evidenced by the case
$q^{2}>m^{2}$.

 \vskip1.20cm \noindent \textbf{4.} --
\emph{A coordinate choice} \`a la
\emph{Brillouin(-Schwarzschild)}. -- For $m^{2}\geq q^{2}$, a
reasonable choice of a new radial coordinate removes the
singularity at $r=0$ of eq. (\ref{eq:one}); indeed, let us perform
the following substitution:

\begin{equation} \label{eq:fourteen}
r \rightarrow r+r_{1} \quad; \quad \left(r_{1}\equiv
m+(m^{2}-q^{2})^{1/2}\right) \quad;
\end{equation}

at $r=0$, the new $\gamma$ is zero, which is its minimal value.
Our manifold is maximally extended and geodesically complete. This
form of the metric is diffeomorphic to the exterior part $r\geq
r_{1}$ of the form (\ref{eq:one}). The known generalization
\emph{\`a la} Kruskal-Szekeres of the $\textrm{d}s^{2}$ of eq.
(\ref{eq:one}) is quite superfluous, and misrepresents the
permanent gravitational field created by $m$ and $q$.

\par For $q=0$, substitution (\ref{eq:fourteen}) becomes $r \rightarrow
r+2m$, \emph{i.e.}, that substitution of the coordinate $r$ of the
\emph{standard} metric of Schwarzschild manifold, which gives the
Brillouin(-Schwarzschild) $\textrm{d}s^{2}$ , see \cite{9}.

\par In a reasonable treatment \emph{there is no room for a real existence of black
holes}. The consideration of the Brillouin(-Schwarzschild) metric
reinforces the conclusion of sects. \textbf{2} and \textbf{3}.

\par A last remark. We think that the choice of the global time
$t=x^{0}$ (Weyl's ``kosmische Zeit'' \cite{2}) as dynamical
evolution parameter for the radial geodesics of Schwarzschild,
Kerr and Rei\ss ner-Weyl-Nordstr\"om  manifolds is the most
appropriate one from the physical point of view \cite{8}. In
particular, it allows a unique treatment of the trajectories of
test-particles and light-rays, while the employment of the proper
time (which is influenced by particle motion) is limited to the
geodesics of the test particles.

% \newpage
\vskip2.00cm
\begin{center}
\noindent \small \emph{\textbf{APPENDIX}}
\end{center} \normalsize

\vskip0.40cm \noindent With the metric of eq. (\ref{eq:one}) the
electrostatic field $F_{01}(r)$ created by the charged gravitating
mass $m$ is simply equal to $\varepsilon/r^{2}$ (``merit'' of the
term $r^{2}\,\textrm{d}\omega^{2}$ in eq. (\ref{eq:one})). As it
was remarked by von Laue \cite{6}, this is an expression ``die
sich von der entsprechenden der Elektrostatik im Euklidischen
dadurch unterscheidet, da\ss {} $r$ nicht den nat\"urlich
gemessenen Abstand vom Ladungstr\"ager bedeutet.''

\par With a different form of the metric we have a
more complicated expression of the e.s. field. For instance, in
Bergmann's metric \cite{7} the e.m. potential $\Phi$ is not given
by $-\varepsilon/r$, but by the following formulae (where
$\varepsilon$ is equal to $-\varepsilon$ of eq. (\ref{eq:one})):

\begin{equation} \label{eq:A1}
\Phi = \frac{\varepsilon}{(q^{2}-m^{2})^{1/2}} \, \cdot \,
\textrm{arccot}\left[ \frac{r-m}{(q^{2}-m^{2})^{1/2}}\right]\quad,
\tag{A.1}
\end{equation}

if $q^{2} \equiv 4 \pi \varepsilon^{2}>m^{2}$; and by

\begin{equation} \label{eq:A2}
\Phi = -\frac{\varepsilon}{2(m^{2}-q^{2})^{1/2}} \, \cdot \, \ln
\left|
\frac{r-m-(m^{2}-q^{2})^{1/2}}{r-m+(m^{2}-q^{2})^{1/2}}\right|
\quad, \tag{A.2}
\end{equation}

if $m^{2}>q^{2}$. When $m^{2}=q^{2}$, we have

\begin{equation} \label{eq:A3}
\Phi = \frac{\varepsilon}{r-m} \quad. \tag{A.3}
\end{equation}

For $\textrm{d}\Phi / \textrm{d}r$ there is a unique formula:

\begin{equation} \label{eq:A4}
\frac{\textrm{d}\Phi}{\textrm{d}r} =
\frac{-\varepsilon}{r^{2}-2\,m\,r+q^{2}} \quad. \tag{A.4}
\end{equation}

It is interesting that at $r=m\pm (m^{2}-q^{2})^{1/2}$ the field
$\textrm{d}\Phi / \textrm{d}r$  and the potential $\Phi$ have
infinite values.

\par Bergmann's formulae make particularly evident that in GR e.m.
field and gravitational field influence each other in a
significant way.

\end{document}